\documentclass[11pt]{article}

\usepackage{latexsym, amssymb, amsmath}

\newcommand{\be}{\begin{equation}}
\newcommand{\ee}{\end{equation}}
\newcommand{\beq}{\begin{eqnarray}}
\newcommand{\eeq}{\end{eqnarray}}

\newtheorem{thm}{Theorem}
\newtheorem{cor}{Corollary}
\newtheorem{lma}{Lemma}

\newtheorem{prop}{Proposition}

\def\n{\nu}
\def\R{\mathbb R}

\def\bg{\bar{g}}
\def\lap{\triangle}
\def\R{\mathbb{R}}

\def\S{\Sigma}
\def\pf{\noindent {\em Proof: }}
\def\vh{\vspace{.2cm}}

\def\vh{\vspace{.2cm}}
\def\p{\prime}
\def\n{\nu}
\def\R{\mathbb R}

\def\bg{\bar{g}}
\def\lap{\triangle}
\def\R{\mathbb{R}}

\def\S{\Sigma}

\def\pf{\noindent{\em Proof: }}

\begin{document}

\title{A Remark on Boundary Effects in Static Vacuum  Initial Data sets}
\date{April 18, 2005}
\author{Pengzi Miao  \footnote{Department of Mathematics, University of
California, Santa Barbara, CA 93106, USA, pengzim@math.ucsb.edu}
}
\maketitle

\begin{abstract}
Let $(M, g)$ be an asymptotically flat static vacuum 
initial data set with non-empty compact boundary $\S$.
We prove that $(M, g)$ is isometric to 
a spacelike slice 
$(\mathbb{R}^3 \setminus B_{\frac{m}{2}}(0), 
(1 + \frac{m}{2|x|} )^4 \delta_{ij})$ 
of a Schwarzschild spacetime under the mere assumption that $\S$ has
zero mean curvature, hence generalizing a classic result of 
Bunting and Masood-ul-Alam. In the case that $\S$ has constant positive mean 
curvature and satisfies a stability condition, we derive an upper bound of 
the ADM mass of $(M, g)$ in terms of the area and mean curvature of $\S$. 
Our discussion is motivated by Bartnik's quasi-local mass definition.
\end{abstract}

\section{Introduction}
Throughout this paper we let $(M, g)$ be an asymptotically flat three 
dimensional Riemannian manifold that has one end and non-empty smooth 
compact boundary $\S$. Following \cite{Corvino}, we say $(M, g)$ is 
{\em static} if  there exists a function $u$, called the 
{\em static potential} of $(M, g)$, satisfying 
$u \rightarrow 1$ at the end of $(M, g)$ and
\be \label{stateq}
\left\{
\begin{array}{ccc}
 u Ric(g) & = &  D^2 u  \\
\lap u  & = & 0,
\end{array}
\right.
\ee
where $D^2 u$ is the Hessian of $u$ and $Ric(g)$ is the Ricci curvature of
$g$. It is well known that if $(M, g)$ and $u$ satisfy (\ref{stateq}), the
asymptotically flat spacetime metric $\bg = -u^2 dt^2 + g$ solves the
Vacuum Einstein Equation on $M \setminus u^{-1}(0) \times \R$, 
where $u^{-1}(0)$ is the zero-set of $u$. It is also known that 
$u^{-1}(0)$, if non-empty, is an embedded totally geodesic 
hypersurface in $M$ \cite{Corvino}. As a result, (\ref{stateq}) implies
that $(M, g)$ has zero scalar curvature.

A classic result of Bunting and Masood-ul-Alam \cite{B-M} states that,
if  $(M, g)$ is static and $u = 0$ on $\Sigma$, $(M, g)$ is 
isometric to a spacelike slice 
$(\mathbb{R}^3 \setminus B_{\frac{m}{2}}(0), 
(1 + \frac{m}{2|x|} )^4 \delta_{ij})$ 
of a Schwarzschild spacetime with positive mass $m$. In particular, 
$\Sigma$ is a connected two sphere.

The condition $u = 0$ on $\S$ is a natural assumption as it corresponds to
that the static killing vector field $\partial_t$ of $\bg$
vanishes at $\Sigma$. On the other hand, from a point of view of 
quasi-local mass question, the mean curvature of $\Sigma$, denoted
by $H_\Sigma$, and the induced metric on $\Sigma$, denoted by $g|_\Sigma$,
also represent important boundary condition. In \cite{Bartnik_local}, Bartnik
proposed his quasi-local mass definition and $H_\Sigma$ and $g|_\Sigma$ 
constitute the geometric boundary constraint in his static metric extension
conjecture \cite{Bartnik_energy} \cite{Miao_static}. Thus,
it is natural to ask whether Bunting and Masood-ul-Alam's result \cite{B-M}
holds under the mere assumption $H_\S = 0$. We give a positive answer
to this question.

\begin{thm} \label{thmone}
Let $(M, g)$ be an asymptotically flat manifold with non-empty smooth compact
boundary $\S$. If $(M, g)$ is static and $\S$ has zero mean curvature, then
 $(M, g)$ is isometric to a spacelike slice 
$(\mathbb{R}^3 \setminus B_{\frac{m}{2}}(0), 
(1 + \frac{m}{2|x|} )^4 \delta_{ij})$ 
of a Schwarzschild spacetime with positive mass $m$. In particular, 
$\Sigma$ is a connected two sphere.
\end{thm}

An immediate corollary of Theorem \ref{thmone} is a non-existence result
on horizons(stable minimal two spheres) in a static and asymptotically
flat manifold whose boundary has non-negative mean curvature. Our sign
convention on mean curvature gives $H_{\partial B_1(0)} = 2$, where
$\partial B_1(0)$ is the boundary of the Euclidean exterior region 
$\mathbb{R}^3 \setminus B_1(0)$.

\begin{cor} \label{nohorizon}
Let $(M, g)$ be an asymptotically flat manifold with non-empty smooth compact
boundary $\S$. If $(M, g)$ is static and $\S$ has non-negative mean curvature,
then there is no horizon (allowed to have multiple components) in $M$ 
enclosing $\Sigma$.
\end{cor}

We end this paper by applying the idea of the proof of Theorem \ref{thmone} 
to the case where $\S$ has constant positive mean curvature. As a result, 
we derive an upper bound of the ADM mass of $(M, g)$ in terms of the 
area and mean curvature of $\S$(see Section \ref{massbd}).

\section{Proof of the Theorem}
We assume that $g$ has sufficient boundary regularity, say
$g$ is $C^3$ on the closure of $M$, then $u$ is $C^2$ on the closure
of $M$ \cite{Corvino}. We first prove Theorem \ref{thmone} under an 
additional assumption
that $u$ is entirely positive in $M$.

\begin{prop} \label{propone}
Assume that $(M, g)$ is asymptotically flat and static. Suppose 
that $(M, g)$ admits a positive static potential $u$. If 
$\S$ has zero mean curvature, then $(M, g)$ is isometric 
to a spacelike slice $(\mathbb{R}^3 \setminus B_{\frac{m}{2}}(0), 
(1 + \frac{m}{2|x|} )^4 \delta_{ij})$ 
of a Schwarzschild spacetime with positive mass $m$.
\end{prop}

\pf 
The asymptotic flatness of $g$ and the equation $\lap u = 0$ guarantee that
$u$ has an asymptotic expansion at $\infty$
\be \label{uexpan}
u = 1 - \frac{m}{|x|} + O(|x|^{-2}) ,
\ee
where $m$ is some constant. It is a non-trivial fact of (\ref{stateq}) 
that $m$ is indeed the ADM mass of
$(M, g)$ \cite{B-M}. We now consider the Green function $G$ at $\infty$
 defined by
\be
\left\{
\begin{array}{ccl}
\lap G & = & 0 \ \ \ in \ M \\
G & \rightarrow & 1 \ \ \ at \ \ \infty \\
G & = & 0 \ \  \ at  \ \ \Sigma  .
\end{array}
\right.
\ee
$G$ has its own asymptotic expansion
\be
G = 1 - \frac{A}{|x|} + O(|x|^{-2}).
\ee
Since $u \geq 0$ on $\S$, it follows from the maximum principle that
$u \geq G$ on $M$, hence $m \leq A$.
On the other hand, it is proved by Bray (Theorem 9 on page 206 in
\cite{Bray}) that $m \geq A$ for any asymptotically flat manifold $M$ whose 
scalar curvature is non-negative and boundary mean curvature is zero, 
furthermore the equality holds if and only if $(M, g)$ is 
isometric to a manifold $(\mathbb{R}^3 \setminus B_{\frac{m}{2}}(0), 
(1 + \frac{m}{2|x|} )^4 \delta_{ij})$.
Therefore, we have $m = A$ and Proposition \ref{propone} follows from 
Bray's theorem.

\vh

Next, we show that Theorem \ref{thmone} holds if $\S$ consists of 
{\em outermost} minimal surfaces of $(M, g)$.
A minimal surface is called {\em outermost} \cite{Bray} if it is not
contained entirely inside another minimal surface. 
Here ``contained entirely inside'' is defined with 
respect to the end of $(M, g)$. 

\begin{prop} \label{proptwo}
Assume that $(M, g)$ is asymptotically flat and static.
If $\S$ consists of outermost minimal surfaces of $(M, g)$,
then $(M, g)$ is isometric to a spacelike slice 
$(\mathbb{R}^3 \setminus B_{\frac{m}{2}}(0), 
(1 + \frac{m}{2|x|} )^4 \delta_{ij})$ 
of a Schwarzschild spacetime with positive mass $m$.
\end{prop}

To prove Proposition \ref{proptwo}, we need a lemma concerning 
the equation of $u$ when restricted to a hypersurface.
For a given hypersurface $\S \subset M$, we let $\lap_\S$ denote 
the Laplacian operator of the induced metric on $\S$, $\n$ denote 
the unit normal vector at $\S$ pointing to $\infty$, $A_\S$ denote 
the second fundamental form of $\S$ and $K_\S$ be the 
Gaussian curvature of $(\S, g|_\S)$.

\begin{lma} \label{lmaone}
Assume that $(M, g)$ is static and $u$ is a static potential. Suppose that 
$\S \subset M$ is a smooth compact hypersurface. 
Then the restriction of $u$ to 
$\S$ satisfies
\be \label{ures}
\lap_\Sigma u + H_\Sigma \frac{\partial u}{\partial \n} + 
(  \frac{1}{2} H^2_{\Sigma} - \frac{1}{2} |A_{\Sigma}|^2  - K_\S) u = 0 .
\ee
\end{lma}

\pf Let $U$ be a Gaussian tubular neighborhood of $\S$ in $M$ 
such that $U$ is diffeomorphic to $\S \times (-\epsilon, \epsilon)$ 
and $g|_U$ has the form $ g = g_t + dt^2 $,
where $t$ is the coordinate of $( - \epsilon, \epsilon)$ and $g_t$ is
the induced metric on the slice $\Sigma \times \{t\}$. We arrange the
direction of $\partial_t$ so that $\partial_t$ points to $\infty$.
At $\S$, we have
\be \label{f}
\lap u  =  \lap_{\Sigma} u + H_\Sigma \frac{\partial u}{\partial t}
+ \frac{\partial ^2 u}{\partial t^2},
\ee
where 
$\frac{\partial ^2 u}{\partial t^2}$ agrees with $D^2u(\partial_t, \partial_t)$
because $\partial_t$ is the velocity vector of a geodesic.
Hence it follows from (\ref{stateq}) that
\be \label{rv}
\lap_\S u + H_\S \frac{\partial u}{\partial \n} + Ric(\n, \n) u = 0 .
\ee
Applying the Gauss equation and using the fact $g$ has zero scalar 
curvature, we have
\be \label{ric}
Ric(\nu, \nu) = \frac{1}{2}H^2_\S - \frac{1}{2} |A_\S|^2 - K_\S.
\ee
Lemma \ref{lmaone} follows from (\ref{rv}) and (\ref{ric}).

\vh

\noindent {\em Proof of Proposition \ref{proptwo}: }
The assumption that $\S$ consists of outermost minimal surfaces implies that 
$\S$ is area {\em outer minimizing} \cite{Bray} \cite{IMF}, i.e.
there is no other closed surface enclosing $\S$ which has less area 
than $\S$. In particular, $\S$ is stable with respect to the second 
variation of area inside $(M, g)$, hence we have
\be \label{stability}
\int_\S | \nabla_\S f|^2 - (Ric(\nu, \nu) + |A_\S|^2) f^2 \geq 0 
\ee
for any $f \geq 0$ on $\S$. Now by a general fact that, for any fixed $h$,
\be
 \inf_{f > 0} \left\{
 \frac{ \int_{\S} |\nabla_{\S} f|^2 - h f^2}
{\int_{\S} |f|^2} \right\} = 
\inf_{f \in W^{1,2}(\S)} \left\{
\frac{ \int_{\S} |\nabla_{\S} f|^2 - h f^2}
{\int_{\S} |f|^2} \right\},
\ee
we see (\ref{stability}) holds without 
requiring $f \geq 0$.
In particular, we can choose $f = u$ to have
\be \label{fequ}
\int_\S | \nabla_\S u|^2 - (Ric(\nu, \nu) + |A_\S|^2) u^2 \geq 0 .
\ee
On the other hand, it follows from Lemma \ref{lmaone} and the
assumption $H_\S =0$ that
\be
\lap_\Sigma u +  Ric(\nu, \nu) u = 0 .
\ee
Multiplying it by $u$ and integrating by parts, we have
\be \label{a1}
\int_\S \left[ | \nabla_\S u|^2 - ( Ric(\nu, \nu) + |A_\S|^2 )u^2 \right]
+ |A_\S|^2 u^2 = 0.
\ee
Hence, (\ref{fequ}) and (\ref{a1}) imply that 
\be \label{a2}
\int_\S | \nabla_\S u|^2 - ( Ric(\nu, \nu) + |A_\S|^2 )u^2  = 0.
\ee
It follows from (\ref{stability}) and (\ref{a2}) that,
on each connected component of $\S$, either $u$ is identically zero 
or $u$ is the first eigenfunction with eigenvalue $0$ of
the operator 
$\lap_\S - (Ric(\nu, \nu) + |A_\S|^2) ,$
in which case $u$ must not change sign.
Suppose $u < 0$ on some component of $\S$, we consider
\be
M_+ = \{ x \in M \ | \ u(x) > 0 \}.
\ee
It follows from the fact  $u^{-1}(0)$ 
is an embedded totally geodesic hypersurface on which 
$\frac{\partial u}{\partial \nu}$ is
a non-zero constant that $M_+$ has smooth compact boundary $\partial M_+$ 
on which $u \geq 0$. Let $M^\infty_+$ be the connected 
component of $M_+$ that contains the asymptotic flat end of $M$, it 
follows from Proposition \ref{propone} that $(M^\infty_+, g)$ is 
isometric to a Schwarzschild spacetime slice
$(\mathbb{R}^3 \setminus B_{\frac{m}{2}}(0), 
(1 + \frac{m}{2|x|} )^4 \delta_{ij})$.
In particular, $\partial M^\infty_+$ is the unique connected outermost
horizon of $(M, g)$ on which $u=0$. 
Therefore, $\S$ must agree with $\partial M^\infty_+$
by the assumption that $\S$ consists of outermost minimal surfaces 
of $(M, g)$. Hence, $u =0$ on $\S$ which contradicts to the assumption 
that $u<0$ on some component of $\S$. 
We conclude that $u$ must be non-negative on $\S$ and Proposition
\ref{proptwo} follows from Proposition \ref{propone}.

\vh

We now can finish the proof of Theorem \ref{thmone}.

\vh

\noindent {\em Proof of Theorem \ref{thmone}:} The assumption that 
$\S$ is a minimal surface implies that the outermost minimal surface
of $(M, g)$, denoted by $\S^\p$, always exists in $M$ \cite{Bray} \cite{IMF}
($\S^\p$ may have multiple components).
Let $M^\infty$ be the region of $M$ outside $\S^\p$, Proposition 
\ref{proptwo} implies that $(M^\infty, g)$ is isometric to a
Schwarzschild spacetime slice $(\mathbb{R}^3 \setminus B_{\frac{m}{2}}(0), 
(1 + \frac{m}{2|x|} )^4 \delta_{ij})$. 
We now resort to a recent result of Chru{\'s}ciel \cite{Chrusciel_anal} 
on the analyticity of static vacuum metrics at non-degenerate horizons. 
By the section 4 in \cite{Chrusciel_anal}, $(M, g)$ admits a 
global analytic atlas (even across $u^{-1}(0)$) with respect
to which $u$ and $g$ are both analytic. Since the unique analytic 
continuation of the manifold
$(\mathbb{R}^3 \setminus B_{\frac{m}{2}}(0), 
(1 + \frac{m}{2|x|} )^4 \delta_{ij})$ is the whole Schwarzschild spacetime 
slice $M^S = (\R^3 \setminus \{0\}, (1 + \frac{m}{2|x|})^4 \delta_{ij})$,  
we see that $(M, g)$ is isometric to a region $D$ of $M^S$, where
$D$ contains the entire upper half end
$(\mathbb{R}^3 \setminus B_{\frac{m}{2}}(0), 
(1 + \frac{m}{2|x|} )^4 \delta_{ij})$
and $\partial D$ is a minimal surface.
However, $M^S$ contains no minimal surface other than its neck at 
$\{|x| = \frac{m}{2} \}$, we conclude that $\S$ agrees with $\S^\p$.
Theorem \ref{thmone} is proved.

\vh

A similar argument also gives a proof of Corollary \ref{nohorizon}.

\vh

\noindent {\em Proof of Corollary \ref{nohorizon}: }
Suppose that $(M, g)$ admits a horizon $\S^\p$ enclosing $\S$ ($\S^\p$ 
may have multiple components). Let $M^\infty$ be the region of $M$ outside
$\S^\p$, Theorem \ref{thmone} implies that $(M^\infty, g)$ is isometric 
to a Schwarzschild spacetime slice
$(\mathbb{R}^3 \setminus B_{\frac{m}{2}}(0), 
(1 + \frac{m}{2|x|} )^4 \delta_{ij})$. 
By the same reasoning as in the proof of Theorem \ref{thmone},
we know that $(M, g)$ is isometric to a region $D$ in a whole 
Schwarzschild spacetime slice
$M^S = (\R^3 \setminus \{0\}, (1 + \frac{m}{2|x|})^4 \delta_{ij})$, 
where $\partial D$ is a closed surface in the lower half end
$(B_\frac{m}{2}(0) \setminus \{0\}, (1 + \frac{m}{2|x|})^4 \delta_{ij})$
of $M^S$. Since $H_\S \geq 0$, the mean curvature of
$\partial D$ computed with respect to the unit normal vector pointing towards
the horizon neck $\{ |x| = \frac{m}{2} \}$ is non-negative.  However,
$(B_\frac{m}{2}(0) \setminus \{0\}, (1 + \frac{m}{2|x|})^4 \delta_{ij})$ is
foliated by a family of constant negative mean curvature surfaces
$\{ |x| = r \}$, the maximum principle implies that $\partial D$
does not exist. Hence, there is no horizon in $(M, g)$ enclosing $\S$ 
and Corollary \ref{nohorizon} is proved.

\section{An upper bound of the ADM mass} \label{massbd}
In \cite{Bartnik_local} Bartnik proposed his quasi-local mass definition
for a compact region $(\Omega, g_\Omega)$ isometrically contained in an 
asymptotically flat manifold with non-negative scalar curvature. He further
conjectured that there exists a static and asymptotically flat manifold 
$(M_b, g_b)$ with boundary, depending only on the boundary data 
$H_{\partial \Omega}$ and $(\partial \Omega, {g_\Omega}|_{\partial \Omega})$,
such that the ADM mass of $(M_b, g_b)$ achieves the Bartnik's
quasi-local mass of $(\Omega, g_\Omega)$. Therefore, as a step to understand 
Bartnik's quasi-local mass of a boundary surface, it is interesting to 
estimate the ADM mass of a given static and asymptotically flat manifold 
$(M, g)$ in terms of its boundary data $H_\S$ and $(\S, g|_\S)$.

We consider the simplest case where $H_\S = H_0$, a positive constant. 
Motivated by the role played by stable minimal surfaces in the proof of 
Theorem \ref{thmone}, we assume $\S$ satisfies an additional stability 
assumption. For a constant mean curvature surface $\S^\p$, we say 
$\S^\p$ is {\em mean-stable} if
\be \label{mstb}
\int_{\S^\p} | \nabla_{\S^\p} 
\phi|^2 - (Ric(\nu, \nu) + | A_{\S^\p} |^2 ) \phi^2 \geq 0
\ee
for any $\phi$ satisfying $\int_{\S^\p} \phi = 0$.
A result of Huisken and Yau \cite{H_Y} states that, 
for any asymptotically flat
manifold $(M^3, g)$ whose mass is strictly positive, 
there is a unique foliation of mean-stable constant mean curvature spheres 
in the asymptotic region.

\begin{prop} \label{upmassbd}
Let $(M, g)$ be an asymptotically flat and static manifold with connected
boundary $\S$. Assume that $H_\S = H_0$ and $K_\S \geq \frac{1}{4} H_0^2$. 
Then, if $\S$ is mean-stable,  
\be
m \leq  4 \sqrt{\frac{16 \pi}{ H_0^2 |\S|}} m_H(\S),
\ee
where $m$ is the ADM mass of $(M, g)$, $H_0$ is a positive constant, 
$|\S|$ is the area of $\S$ and $m_H(\S)$ is the Hawking quasi-local mass 
of $\S$. The equality holds if and only if $(M, g)$ 
is Euclidean.
\end{prop}

\pf
We first show that the assumption $K_\S \geq \frac{1}{4}H^2_\S $
and $H_\S > 0$ guarantees that $ 0 < u < 1$ in $M$ unless $g$ is flat.
Applying Lemma \ref{lmaone} to $\S = \partial M$,
we have
\be \label{bkhteq}
-  H_\Sigma \frac{\partial u}{\partial \n} = \lap_\Sigma u +
( \frac{1}{2} H^2_{\Sigma} - \frac{1}{2} |A_{\Sigma}|^2 - K_\S ) u .
\ee
Let $p \in \S$ such that $u(p) = \min_\S u$. Suppose $u(p) \leq 0$,
then $u(p) = \min_M u$ because $\lap u = 0$ and
$u \rightarrow 1$ at $\infty$. Hence, 
$\frac{\partial u}{\partial \n}(p) > 0$ by the Hopf strong maximum
principle. Therefore,
$H_\S(p) \frac{\partial u}{\partial \n}(p) > 0$ by the assumption 
$H_\S > 0$.
On the other hand,
\be \label{h}
\frac{1}{2} H^2_{\Sigma}(p) - \frac{1}{2} |A_{\Sigma}|^2(p)
- K_\S(p) \leq \frac{1}{4} H^2_\S(p)  - K_\S(p) \leq 0
\ee
by the assumption $\frac{1}{4} H^2_\S \leq K_\S$, and
$ \lap_\S u(p) \geq 0 $ by the maximum principle.
Therefore, we get a contradiction to (\ref{bkhteq}).
Hence, $\min_\S u > 0$. A similar argument shows that $\max_\S u < 1$
unless $u \equiv 1$, in which case $(M, g)$ is flat. It then follows from 
the maximum principle that $0 < u < 1$ in $M$ unless $u \equiv 1$.

Next, we define $ v = \log{u}$, it follows from
(\ref{ures}) that
\be \label{c}
\lap_\Sigma v + |\nabla_\Sigma v|^2 + 
H_\Sigma \frac{\partial v}{\partial \n} = 
\frac{1}{2} ( 2 K_\S - H^2_\Sigma + |A_\Sigma|^2  ) .
\ee
On the other hand,
\be
\int_{S_\infty} \frac{\partial v }{\partial \n} = \int_{S_\infty} 
\frac{\partial u}{\partial \n} = 4 \pi m
\ee
and
\be \label{b}
\lap v + |\nabla v|^2 = \frac{1}{u} \lap u = 0.
\ee
Integration by parts, we have
\be \label{ab}
4 \pi m + \int_M |\nabla v|^2 = \int_\Sigma \frac{\partial v}{\partial \n} .
\ee
Integrating  (\ref{c}) on $\S$ and applying $H_\S = H_0$ and (\ref{ab}),
we have
\be \label{dd}
\int_\Sigma |\nabla_\Sigma v|^2 + 
 H_\S \int_M |\nabla v|^2 + 4 \pi m H_\S = 
\frac{1}{2} \int_\Sigma ( |A_\S|^2 - H^2_\S ) + 4 \pi,
\ee
where we used $\int_\S K_\S = 4 \pi$ by the Gauss-Bonnet theorem and 
the fact that $K_\S > 0$.
We now apply the mean-stable condition to get a $L^2$ estimate of $|A_\S|$.
We follow an idea in \cite{C_Y} and choose $\psi$ to be a conformal map of 
degree $1$ which maps $(\S, g|_\S)$ onto the standard sphere 
$S^2 \subset \mathbb{R}^3$. Using the conformal group of $S^2$, we can arrange
that each component $\psi_i$ of $\psi$, $i = 1, 2, 3$, satisfies
$ \int_\S \psi_i = 0$.
On the other hand, the Dirichlet integral is conformal invariant in 
dimension $2$, so  
$$ \int_\S | \nabla_\S  \psi_i |^2 = \int_{S^2} | \nabla_{S^2} x_i |^2 =
\frac{ 8 \pi}{3}. $$
Applying the mean-stability condition (\ref{mstb}) to $\psi_i$ and 
summing over $i$, we get
\be \label{bbb}
8 \pi \geq \int_\S ( Ric(\nu, \nu ) + |A_\S|^2) .
\ee
It follows from the Gauss equation and the fact $g$ has zero scalar curvature
 that
\be
Ric(\nu, \nu) + |A_\S|^2 = \frac{1}{2} H_0^2 + \frac{1}{2}|A_\S|^2 - K_\S.
\ee
Hence, (\ref{bbb}) implies 
\be \label{ltwoa}
12 \pi - \frac{1}{2} \int_\S H_0^2 \geq \frac{1}{2} 
\int_\S |A_\S|^2
\ee
by the Gauss-Bonnet theorem. It follows from (\ref{dd}) that
\be \label{d}
\int_\Sigma |\nabla_\Sigma v|^2  +  H_0 \int_M |\nabla v|^2 + 4 \pi m H_0
\leq 16 \pi - \int_\S H_0^2 . 
\ee
Hence,
\be \label{xyz}
\frac{1}{4 \pi H_0} 
\left[ \int_\Sigma |\nabla_\Sigma v|^2  + H_0 \int_M |\nabla v|^2  \right]
+  m \leq  4 \sqrt{\frac{16 \pi}{ H_0^2 |\S|}} m_H(\S),
\ee
where 
\be
 m_H(\S) = \sqrt{\frac{|\S|}{16 \pi}} \left(1 - \frac{1}{16 \pi}
\int_\S H^2 \right)
\ee
is the Hawking quasi-local mass of $\S$. Proposition \ref{upmassbd}
follows from (\ref{xyz}).

\begin{cor} \label{euclidean}
Let $(M, g)$ be an asymptotically flat and static manifold with 
boundary $\S$. Assume that $(\S, g|_\S)$ is isometric to the standard 
unit sphere $S^2 \subset \mathbb{R}^3$ and $\S$ has constant mean 
curvature $2$. If $\S$ is mean-stable, then $(M^3, g)$ is isometric to 
$\mathbb{R}^3 \setminus B_1(0)$. 
\end{cor}

\pf The boundary assumption implies that $m_H(\S) = 0$. Hence, Proposition
\ref{upmassbd} implies that $m \leq 0$. On the other hand, we can glue
the Euclidean ball $B_1(0)$ and $(M, g)$ along the boundary and the 
generalized positive mass theorem in \cite{Miao_PMT} implies that $m \geq 0$. 
Therefore, $m=0$ and $(M, g)$ is isometric to the Euclidean exterior region 
$\mathbb{R}^3 \setminus B_1(0)$ by the theorem in \cite{Miao_PMT}.

\vh

\noindent {\em Remark }
{One can also prove Corollary \ref{euclidean} by showing that 
(\ref{ltwoa}) implies $A_\S = g|_\S$ and proving $u \equiv 1$
in a way similar to the derivation of $0 < u < 1$ in 
Proposition \ref{upmassbd}.
We choose the above proof to demonstrate the expectation that a static metric
might be the minimal mass metric, hence minimizes the ADM mass.}

\vh

\noindent {\bf Acknowledgement}
The author is grateful to P Chru{\'s}ciel for explaining the analyticity
of static vacuum metrics at non-degenerate horizons.


\end{document}